\documentclass[aps,pra,twocolumn,twoside,superscriptaddress,notitlepage]{revtex4-2}
\usepackage{tikz}
\usetikzlibrary{calc}
\usepackage{eufrak}
\usepackage{enumitem}
\usepackage{mathtools}
\usepackage{graphicx}
\usetikzlibrary{positioning}
\usepackage{bm}
\newcommand{\bematrix}{\left(\begin{matrix}}
\newcommand{\ematrix}{\end{matrix}\right)}
\usepackage{ulem} 
\usepackage[caption=false]{subfig}
\usepackage{dcolumn}
\usepackage{amsmath,amssymb}
\usepackage{bm}
\usepackage{bbm}
\usepackage{overpic}
\usepackage{latexsym}
\usepackage{color}
\usepackage[english]{babel}
\usepackage{latexsym}
\usepackage{psfrag,graphicx}
\usepackage{epsf}
\usepackage{amsmath}
\usepackage{amssymb}
\usepackage{amsfonts}
\usepackage{natbib}

\usepackage{appendix}
\usepackage{verbatim}
\usepackage{enumitem}
\usepackage{dsfont}

\usepackage{tikz}

\definecolor{mygrey}{gray}{0.35}
\definecolor{myblue}{rgb}{0.2,0.2,0.8}
\definecolor{myzard}{cmyk}{0,0,0.05,0}
\definecolor{mywhite}{rgb}{1,1,1}
\definecolor{myred}{rgb}{0.9,0.1,0.}
\usepackage[colorlinks=true,citecolor=myblue,linkcolor=myblue,urlcolor=myblue]{hyperref}

\usepackage[makeroom]{cancel}

\newtheorem{theorem}{Theorem}

\newtheorem{definition}[theorem]{Definition}

\newenvironment{proof-of}[1]{\medskip\noindent\textbf{Proof of {#1}.}}{\hfill$\blacksquare$\medskip}
\newcommand{\ket}[1]{\left\vert#1\right\rangle}

\newcommand{\ketbra}[2]{\ensuremath{| #1 \rangle\!\langle #2 |}}
\newcommand{\norm}[1]{\left\lVert#1\right\rVert}

\begin{document}

\title[ATI]{Entanglement manipulation through multicore fibres}
\author{Carlo Marconi}
\affiliation{Istituto Nazionale di Ottica del Consiglio Nazionale delle Ricerche (CNR-INO), 50125 Firenze, Italy}

\author{Elena Fanella}
\affiliation{Istituto Nazionale di Ottica del Consiglio Nazionale delle Ricerche (CNR-INO), 50125 Firenze, Italy}
\affiliation{Università degli Studi di Napoli Federico II, Napoli, Italy}

\author{Davide Bacco}
\affiliation{Department of Physics and Astronomy, University of Florence, 50019, Firenze, Italy}

\author{Alessandro Zavatta}
\affiliation{Istituto Nazionale di Ottica del Consiglio Nazionale delle Ricerche (CNR-INO), 50125 Firenze, Italy}

\begin{abstract}
\noindent Multicore fibres are recently gaining considerable attention in the context of quantum communication tasks, where their capability to transmit multiple quantum states along different cores of the same channel make them a promising candidate for the implementation of scalable quantum networks. Here, we show that multicore fibres can be effectively used not only for the scope of communication but also for the generation of entangled states. By exploiting the formalism of completely positive trace preserving maps, we describe the action of a multicore fibre as a quantum channel and propose a protocol to implement bound entangled states of two qudits. Notably, the presence of crosstalk among the cores of the fibre is fundamental for the generation of such states. 
\end{abstract}

\maketitle

\section{Introduction}
The ability to manipulate and faithfully transmit quantum states between two or more parties is an essential prerequisite for any quantum communication scheme. This is particularly evident in the recent quest for a quantum internet, which demands the implementation of networks with superior transmission rates, resilience to noise, and seamless integration with current telecommunication setups \cite{wehner2018quantum,azuma2023quantum}. To address these points, one potential strategy involves encoding information in qudits, that is, quantum systems with a Hilbert space of dimension $d>2$ \cite{cozzolino2019high, ecker2019overcoming}. Indeed, it has been shown that qudits offer many notable advantages such as an increased information capacity \cite{barreiro2008beating,dixon2012quantum,hu2018beating}, as well as a greater robustness to environmental noise \cite{sheridan2010security,cerf2002security}, among others. In the context of quantum optics, the physical realisation of a qudit can be achieved through various photonic degrees of freedom, which can be utilized individually or in combination. These include the orbital angular momentum of light \cite{cozzolino2019orbital}, time-energy/time-bin encoding \cite{islam2017provably}, frequency \cite{kues2017chip} and path encoding \cite{krenn2017entanglement}. This latter technique, in particular, offers better performances in terms of scalability when implemented on integrated circuits, although ensuring a reliable transmission of quantum states remains a challenge. 
Multicore fibres (MCFs) \cite{saitoh2016multicore,hayashi2019field}, which contain multiple cores within the same structure, offer a promising solution for improving the transmission of quantum states, being characterised by low crosstalk among different cores \cite{bacco2021characterization}. For this reason, such devices have been used, so far, mostly in the context of quantum communication protocols such as quantum key distribution \cite{da2019stable,da2021path,zahidy2024practical}. Recently, it was shown how MCFs can be used not only for quantum communication purposes but also for entanglement generation \cite{gomez2021multidimensional}. Here, we follow this latter approach and propose a scheme where we use MCFs to manipulate an initial maximally entangled state, resulting in the creation of a new family of mixed entangled states. Remarkably, adjusting the crosstalk parameters of the
MCF, our method allows for the generation of \textit{bound entangled} states \cite{horodecki1998mixed}, which have proven to be relevant in several applications, ranging from quantum metrology \cite{toth2018quantum,toth2020activating} to quantum key distribution \cite{horodecki2005secure}. While the realisation of bound entangled states has already been reported in several experiments \cite{hyllus2004generation,lavoie2010experimental,hiesmayr2013complementarity}, the adopted setups are typically hard to implement. Our approach, in contrast, relies only on the propagation of one subsystem of a maximally entangled state through a single MCF, thus greatly simplifying both the experimental setup and the subsequent certification of entanglement in the output state. This paper is structured as follows: in Section \ref{sec:math} we briefly review the main tools needed for our analysis, recalling the concept of quantum channels and presenting the families of entangled states that will be central to our investigation; in Section \ref{sec:res} we discuss our results, describing the action of a MCF as a quantum channel and presenting a protocol to create the desired entangled states, with a special emphasis on the methods needed to certify the presence of bound entanglement. Finally, we conclude summarising our findings and discussing some open questions.

\section{Mathematical background}
\label{sec:math}

\subsection{Quantum channels}

\noindent Let $\mathcal{H}$ be a finite dimensional Hilbert space and $\mathcal{B}(\mathcal{H})$ denote the space of bounded linear operators on $\mathcal{H}$. The state of a quantum system is described by an operator $\rho \in \mathcal{B}(\mathcal{H})$ such that $\rho \succeq 0$ and $\mbox{Tr}(\rho) = 1$. Any physical transformation on a state is described by a quantum channel, i.e., a linear completely positive trace-preserving (CPTP) map $\mathcal{E}:\mathcal{B}(\mathcal{H}_{A}) \rightarrow \mathcal{B}(\mathcal{H}_{B})$. The properties of a quantum channel $\mathcal{E}$ can be conveniently described in terms of an associated operator $\mathcal{J}_{\mathcal{E}} \in \mathcal{B}(\mathcal{H}_{A} \otimes \mathcal{H}_{B})$, as a consequence of the celebrated Choi-Jamio\l kowski-Sudarshan (CJS) isomorphism \cite{jamiolkowski1972linear,choi1975completely,sudarshan1961stochastic}:
\begin{theorem}[\cite{jamiolkowski1972linear,choi1975completely,sudarshan1961stochastic}]
\label{th:cj}
Given a map $\mathcal{E}:\mathcal{B}(\mathcal{H}_{A}) \rightarrow \mathcal{B}(\mathcal{H}_{B})$ there exists an
associated operator $\mathcal{J}_{\mathcal{E}} \in \mathcal{B}(\mathcal{H}_{A} \otimes \mathcal{H}_{B} )$ defined as
\begin{align}
    \mathcal{J}_{\mathcal{E}} = (\mathds{1}_{A'} \otimes \mathcal{E})\left( \ketbra{\Psi^{+}}{\Psi^{+}}\right)~,
\end{align}
\noindent where $\ket{\Psi^{+}}= \frac{1}{\sqrt{d}} \sum_{i=0}^{d-1}\ket{ii}$ is a maximally entangled state in $\mathcal{H}_{A} \otimes \mathcal{H}_{A'}$ and
$d = \mbox{dim}(\mathcal{H}_{A'}) = \mbox{dim}(\mathcal{H}_{A}) $.

\noindent Conversely, given an operator $\mathcal{J}_{\mathcal{E}} \in \mathcal{B}(\mathcal{H}_{A} \otimes \mathcal{H}_{B} )$
there exists an associated map $\mathcal{E}:\mathcal{B}(\mathcal{H}_{A}) \rightarrow \mathcal{B}(\mathcal{H}_{B})$ defined as
\begin{align}
\mathcal{E}(\rho_{A}) = \emph{Tr}_{A}[\mathcal{J}_{\mathcal{E}}  (\rho_{A}^{T} \otimes \mathds{1}_{B})]~,
\end{align}
\noindent with $\rho_{A} \in \mathcal{B}(\mathcal{H}_{A})$. 
\end{theorem}

\noindent Remarkably, thanks to Theorem \ref{th:cj}, the properties of a quantum map $\mathcal{E}$ can be investigated by turning to its associated Choi operator, as stated by the following result:
\begin{theorem}[\cite{jamiolkowski1972linear,choi1975completely,sudarshan1961stochastic}]
\label{th:choi}
Let $\mathcal{E}:\mathcal{B}(\mathcal{H}_{A}) \rightarrow \mathcal{B}(\mathcal{H}_{B})$ be a quantum map and $\mathcal{J}_{\mathcal{E}} \in \mathcal{B}(\mathcal{H}_{A} \otimes \mathcal{H}_{B})$ its related operator through the CJS isomorphism. Then,
\begin{align}
\label{cp}
    & \mathcal{E}~\emph{completely positive} \iff \mathcal{J}_{\mathcal{E}}\succeq 0~,\\
\label{tp}    
    &\mathcal{E}~\emph{trace-preserving} \iff \emph{Tr}_{B}(\mathcal{J}_{\mathcal{E}})=\frac{\mathds{1}_{A}}{d_{A}}~,
\end{align}
\end{theorem}
\noindent where $d_{A} = \dim (\mathcal{H}_{A})$.

\subsection{PPT-entangled states}
\noindent We briefly review some basic concepts about entanglement in bipartite systems. Let us start by recalling the definition of a separable state.
\begin{definition}
    A bipartite state $\rho_{AB}\in \mathcal{B}(\mathcal{H}_{A} \otimes \mathcal{H}_{B})$ is said separable if it can be decomposed as
    \begin{equation}
        \rho_{AB} = \sum_{k} \lambda_{k} \rho^{(k)}_{A} \otimes \rho^{(k)}_{B}~,
    \end{equation}
    with $\lambda_{k} \geq 0 , ~\sum_{k} \lambda_{k} = 1$, and where $\rho^{(k)}_{A}, \rho^{(k)}_{B}$ are quantum states on $\mathcal{H}_{A}, \mathcal{H}_{B}$, respectively. If such decomposition does not exist, then the state is said to be entangled.
\end{definition}
\noindent It has been shown that, in general, the so-called separability problem, i.e., deciding whether a quantum state is separable or not, is an NP-hard problem \cite{gurvits2003classical}. Nevertheless, in some cases, there exist criteria which are both necessary and sufficient for separability. Before entering the details, let us first recall the definition of the partial transposition of a bipartite state.
\begin{definition}
Let $\rho_{AB} \in  \mathcal{B}(\mathcal{H}_{A} \otimes \mathcal{H}_{B})$ be a bipartite state, whose representation in some product basis is given by
\begin{equation}
    \rho_{AB} = \sum_{ijkl}\rho_{ij,kl}\ketbra{i}{j} \otimes \ketbra{k}{l}~.
\end{equation}
Then, its partial transposition w.r.t.~$B$, i.e., $\rho_{AB}^{T_{B}}$, is defined as
\begin{equation}
\rho_{AB}^{T_{B}} = \sum_{ijkl} \rho_{ij,lk} \ketbra{i}{j} \otimes \ketbra{k}{l}~,
\end{equation}
and an analogous definition holds true for $\rho_{AB}^{T_{A}}$.
\end{definition}

\noindent A state with a positive partial transposition (PPT) is termed PPT; otherwise, it is said to have a non-positive partial transposition (NPT). In the case of bipartite systems, it has been proven that separable states are necessarily PPT \cite{peres1996separability,horodecki1996separability}. Remarkably, a violation of this condition provides a sufficient condition for bipartite entanglement, i.e., any NPT state is entangled. Although one might be tempted to say that every PPT state is always separable, it has been shown this result holds true only when $d_{AB} = \mbox{dim}(\mathcal{H}_{A} \otimes \mathcal{H}_{B}) \leq 6$. Indeed, already when dealing with two-qutrit systems (i.e., $d_{AB} = 9$), there exist states which, in spite of being PPT, are nevertheless entangled and, for this reason, are called PPT-entangled (PPTES). While the entanglement of NPT states is sometimes referred to as \textit{free}, since it can be used as a resource in various quantum information protocols, PPTES are examples of \textit{bound entangled} states, i.e., states from which it is impossible to extract free entanglement through a distillation protocol \cite{horodecki1998mixed}. Although this feature might suggest that PPTES states are useless for quantum information processing, it has been shown that this is not the case \cite{toth2018quantum,toth2020activating,horodecki2005secure,masanes2006all}. The detection of PPTES is usually cumbersome and typically relies on numerical approaches, such as the semidefinite programming technique presented in \cite{doherty2004complete}. However, when restricting to systems endowed with symmetries, the certification of entanglement can be assessed using analytical techniques. One such example is the so-called realignment criterion, which reads:
\begin{theorem}[\cite{chen2002matrix,rudolph2005further}]
\label{th:realign}
    Let $\rho_{AB} \in \mathcal{B}(\mathcal{H}_{A} \otimes \mathcal{H}_{B})$. If $\rho_{AB}$ is separable, then the matrix $\mathcal{R}(\rho_{AB})$ with elements
    \begin{equation}
        \mathcal{R}(\rho_{AB})_{m \mu, n \nu} = (\rho_{AB})_{mn, \nu \mu}
    \end{equation}
    is such that $\norm{\mathcal{R}(\rho_{AB})}_{\emph{Tr}} \leq 1$, where $\norm{ \cdot }_{\emph{Tr}} = \sum_{k} \sigma_{k}(\cdot)$ is the trace-norm of a matrix, i.e., the sum of its singular values $\{\sigma_{k}( \cdot)\}$.
\end{theorem}
\noindent As a consequence, whenever a state $\rho_{AB}$ violates the realignment criterion, i.e., $\norm{\mathcal{R}(\rho_{AB})}_{\mbox{Tr}} > 1$, it is enough to conclude that $\rho_{AB}$ is entangled.

\subsection{Conjugate local diagonal unitary invariant states}

\noindent Let us consider the following state, denoted as $\rho_{(A,B)} \in \mathcal{B}(\mathbb{C}^{d} \otimes \mathbb{C}^{d})$, and formally defined as \cite{singh2021entanglement}:
\begin{align}
\label{CLDUI}
    \rho_{(A,B)} &= \sum_{i,j=0}^{d-1} A_{ij} \ketbra{ij}{ij} + \sum_{i\neq j=0}^{d-1} B_{ij} \ketbra{ii}{jj}~,
\end{align}
\noindent where $A,B$ are $d \times d$ matrices such that $\mbox{diag}(A) = \mbox{diag}(B)$. Notice that, in order for $\rho_{(A,B)}$ to be a quantum state, it is necessary that $A, B$ satisfy $A_{ij} \geq 0~ \forall i,j,~ B\succeq 0$ and $\sum_{ij} A_{ij} = 1$. Moreover, it is easy to see that $\rho_{(A,B)}$ is PPT iff $A_{ij} A_{ji} \geq |B_{ij}|^2~ \forall i,j$. This family of states displays the following symmetry, i.e.,
\begin{equation*}
    \rho_{(A,B)} = (U \otimes U^{\dagger} )\rho_{(A,B)} (U^{\dagger} \otimes U)~, \quad \forall U \in \mathcal{DU}_{d}~,
\end{equation*}
\noindent where $\mathcal{DU}_{d}$ is the group of $d \times d$ diagonal unitary matrices. For this reason, states of the form of Eq.(\ref{CLDUI}) are said \textit{conjugate local diagonal unitary invariant} (CLDUI) states. An exhaustive analysis of CLDUI states has been pursued in \cite{singh2021diagonal}, while a characterisation of their entanglement properties has been tackled in \cite{johnston2018pairwise,singh2021entanglement}. We conclude this section recalling a result regarding the separability of CLDUI states, which will be useful in the following:
\begin{theorem}[\cite{johnston2018pairwise}]
\label{th:realign_AB}
    Let $\rho_{(A,B)}$ be a state of the form of Eq.(\ref{CLDUI}). Then $\rho_{(A,B)}$ satisfies the realignment criterion if and only if 
    \begin{equation*}
        \norm{A}_{1} - \norm{A}_{\emph{Tr}} \leq \norm{B}_{1} - \norm{B}_{\emph{Tr}}~,
    \end{equation*}
    where $\norm{X}_{1}$ is the entrywise 1-norm of $X$, i.e., $\norm{X}_{1} = \sum_{i,j=1}^{d} |X_{ij}|$.
\end{theorem}
As a consequence, a violation of the above condition signals the presence of entanglement in the state $\rho_{(A,B)}$.

\subsection{Diagonal symmetric states}

\noindent In this section we introduce a class of systems whose state remains invariant under any permutation of the parties. Such states, dubbed \textit{symmetric}, describe sets of indistinguishable particles and display an interesting connection with CLDUI states \cite{singh2021diagonal}. In the bipartite case, i.e., $\mathcal{H} = \mathbb{C}^{d} \otimes \mathbb{C}^{d}$, the symmetric subspace $\mathcal{S} \equiv \mathcal{S}(\mathcal{H}) \subset \mathcal{H}$
is spanned by the so called Dicke states, whose expression reads
\begin{equation}
\label{dick2}
\ket{D^{(d)}_{ii}} = \ket{ii}~, \quad \ket{D^{(d)}_{ij}} = \frac{\ket{ij} + \ket{ji}}{\sqrt{2}}~, \quad   i \neq j~,
\end{equation}
where the superscript reminds that $\{\ket{i}\}_{i=0}^{d-1}$ is an orthonormal basis of $\mathbb{C}^{d}$.

\noindent In what follows, we restrict to symmetric states that are diagonal in the Dicke basis, dubbed diagonal symmetric (DS). Such states take the form
 \begin{equation}
 \label{DS}
  \rho_{DS} = \sum_{0 \leq i \leq j < d} p_{ij}\ketbra{D^{(d)}_{ij}}{D^{(d)}_{ij}},
 \end{equation}
with $p_{ij} \geq 0,\; \forall\; i,j$ and $\sum_{ij} p_{ij}=1$.

\noindent A direct calculation shows that the partial transpose of any DS state can be cast as:
 \begin{equation}
 \label{DS:tb}
        \rho_{DS}^{\Gamma}=M_d(\rho_{DS}) \bigoplus_{\substack{{0\leq i \neq j <d}}}\frac{p_{ij}}{2}
    \end{equation}
    where $M_d(\rho_{DS})$ is a $d \times d$ matrix with non-negative entries given by
    \begin{equation}
        M_d(\rho_{DS})=
     \begin{pmatrix}
       p_{00} 	& p_{0 1}/2  & \cdots & p_{0, d-1}/2 \\
       p_{0 1}/2  & p_{11} & \cdots & p_{1, d-1}/2 \\
       \vdots 	& \vdots & \ddots & \vdots \\
       p_{0, d-1}/2 & p_{1, d-1}/2 & \cdots & p_{d-1, d-1} 
    \end{pmatrix} 
    \end{equation}
\noindent Here, $\Gamma$ denotes the partial transposition with respect to either one of the two parties. Interestingly, the partial transposition of a DS state belongs to the class of CLDUI states \cite{singh2021diagonal}, an observation that will be crucial to our analysis.

\noindent In \cite{yu2016separability,tura2018separability} it was shown that, for bipartite DS states, the separability properties of $\rho_{DS}$ can be recast as equivalent properties of its associated matrix $M_{d}(\rho_{DS})$, as expressed by the following Theorem:
\begin{theorem}
[\cite{yu2016separability,tura2018separability}]
\label{th:CAR}
Let $\rho_{DS} \in \mathcal{B}(\mathcal{S} \left( \mathbb{C}^{d} \otimes \mathbb{C}^{d} \right))$ be a DS state. Then,
\begin{align}
\label{th:SEP}
    &\rho_{DS}~\emph{separable} \iff M_{d}(\rho_{DS}) \in \mathcal{CP}_{d}~,\\
\label{th:PPT}
    &\rho_{DS}~ \emph{PPT} \iff M_{d}(\rho_{DS}) \in \mathcal{DNN}_{d}~,\\
\label{th:PPTES}
    &\rho_{DS} ~\emph{PPT-entangled} \iff M_{d}(\rho_{DS}) \in \mathcal{DNN}_{d}\setminus \mathcal{CP}_{d}~,
\end{align}
where $\mathcal{CP}_{d}$ and $\mathcal{DNN}_{d}$ denote, respectively, the cones of \emph{completely positive} and \emph{doubly non-negative} matrices of order $d$, defined as
\begin{align*}
    &\mathcal{CP}_{d} = \{ M = B B^{T}, B_{ij} \geq 0~ \forall i,j\},\\
    &\mathcal{DNN}_{d}=\{M \succeq 0, M_{ij} \geq 0~ \forall i,j \}~.
\end{align*}
\end{theorem}

\noindent Surprisingly, the structure of the aforementioned convex cones depends crucially on the dimension and indeed, it was proved in \cite{diananda_1962,berman2015open} that, for $d<5$, it is $\mathcal{CP}_{d} = \mathcal{DNN}_{d}$. As a consequence, Eqs.(\ref{th:SEP})-(\ref{th:PPT}) imply that, in this case, a DS state is separable if and only if it is PPT. However, when $d \geq 5 $, it is $\mathcal{CP}_{d} \subset \mathcal{DNN}_{d}$, and there exist PPT-entangled DS states $\rho_{DS}$ whose associated matrices satisfy $M_{d}(\rho_{DS}) \in \mathcal{DNN}_{d} \setminus \mathcal{CP}_{d}$ associated to, as expressed by Eq.(\ref{th:PPTES}).

\section{Results}
\label{sec:res}

\subsection{Multicore fibres as quantum channels}

\noindent Let $\{\ket{i}\}_{i=0}^{d-1}$ be an orthonormal basis for the qudit Hilbert space $\mathcal{H}_{in}=\mathbb{C}^{d}$ and let us associate each vector to a different core of the multicore fibre (MCF). A quantum state at the input of the MCF is described by a density operator $\rho_{in} \in \mathcal{B}(\mathcal{H}_{in}),~ \rho_{in}\succeq 0, ~\mbox{Tr}(\rho_{in})=1 $, whose expression in the core basis is given by $\rho_{in} = \sum_{ij} (\rho_{in})_{ij} \ketbra{i}{j}$. The action of a MCF on $\rho_{in}$ can be described in terms of a quantum channel, i.e., a completely-positive trace-preserving (CPTP) map $\mathcal{E}: \mathcal{B}(\mathcal{H}_{in}) \rightarrow \mathcal{B}(\mathcal{H}_{out})$, such that $\rho_{in}$ is mapped into an output state $\rho_{out}$ according to the transformation $\rho_{out} = \mathcal{E}(\rho_{in})$. Although, in principle, the input and output Hilbert spaces might be different, in what follows we restrict to the case where $\mathcal{H}_{in} = \mathcal{H}_{out} \equiv \mathcal{H}$, and the quantum channel $\mathcal{E}$ defines a transformation between states of the same Hilbert space. In order to derive the expression of the quantum channel that corresponds to a MCF, one first requirement comes from the observation that, in the ideal case, this device must faithfully transmit the state of a generic input state $\rho_{in}$. Formally, the action of such a perfect MCF should be described in terms of a quantum channel $\mathcal{E}$ such that
\begin{equation}
\label{stat}
    \mathcal{E} \left( \rho_{in} \right) = \rho_{in}~. 
\end{equation} 

\noindent Any state $\rho_{in}$ that satisfies Eq.(\ref{stat}) is said to be a stationary state or, equivalently, a fixed point for the quantum channel $\mathcal{E}$. Intuitively, in the case of a perfect MCF, Eq.(\ref{stat}) should be satisfied for \textit{any} input state $\rho_{in}$. Nevertheless, it is easy to see that the only CPTP map with an infinite number of stationary states is the identity.
Alternatively, one could restrict to those non-trivial CPTP maps (i.e., different from identity and unitaries) which possess the maximal number of stationary states. Such maps have been recently characterised in \cite{Lewenstein2020} by means of the following theorem:
\begin{theorem}[\cite{Lewenstein2020}]
\label{th:map1}
Let $\{\ket{i}\}_{i=0}^{d-1}$ be an orthonormal basis for the Hilbert space $\mathcal{H}$. Then, any CPTP map $\mathcal{E}: \mathcal{B}(\mathcal{H}) \rightarrow \mathcal{B}(\mathcal{H})$ such that $\mathcal{E}(\ketbra{i}{i}) = \ketbra{i}{i}$ can have up to $d$ stationary states and has the form
\begin{align}
\label{map1}
\mathcal{E}(\rho)=\sum_{i=0}^{d-1} \rho_{i i}\ketbra{i}{i}+ \sum_{i \neq j}^{d-1} \rho_{i j} (1+\alpha_{i j}) \ketbra{i}{j}~,
\end{align}
\noindent where $\alpha_{i j} \in \mathbb{C}, |1+\alpha_{ij}| \leq 1~\forall i,j$. 
\end{theorem}

\noindent Notice that, in order for $\mathcal{E}$ to be a quantum channel, the complex numbers $\alpha_{ij}$ have to be chosen in such a way to ensure complete-positivity, thus leading to a system of highly non-trivial conditions. For instance, in the simpler case $\alpha_{ii} = 0~ \forall i, \alpha_{ij} = \alpha_{ji} \equiv \alpha \in \mathbb{R}~\forall i \neq j$, it is easy to see that this requirement is fulfilled choosing $\alpha \in [-d/(d-1),0]$. 

\noindent From a physical point of view, we can think of the coefficients $\alpha_{ij}$ as the effect of a decoherence process acting on the initial state.  Indeed, it follows from Eq.(\ref{map1}) that the action of $\mathcal{E}$ causes a reduction of the coherences in the input state after each application, i.e., $|\mathcal{E}(\rho_{in})_{ij}|=|(\rho_{in})_{ij}(1+\alpha_{ij})|<|(\rho_{in})_{ij}|$, where the inequality follows from the fact that $|1 + \alpha_{ij}| <1~\forall i\neq j $. As a consequence, any input state which is not diagonal in the chosen basis does not correspond to a stationary state for the channel $\mathcal{E}$. 

\noindent So far, the action of the quantum channel described by Eq.(\ref{map1}) does not take into account the crosstalk among different cores. Such detrimental effect corresponds to the fact that a state which had been initialised in one of the cores can be found, at the output of the MCF, on a different core. Formally, this condition can be described introducing a new map $\mathcal{E}_{C}$ such that
\begin{equation}
    \mathcal{E}_{C}(\ketbra{i}{i}) = \sum_{j=0}^{d-1} p^{(i)}_{j} \ketbra{j}{j}~,
\end{equation}
\noindent where the coefficient $p^{(i)}_{j} \geq 0$ represents the probability that the input state on the $i$-th core is found at the output of the $j$-th core. Hence, a MCF with crosstalk can be described by a quantum map $\mathcal{E}_{C}$ of the form
\begin{equation}
    \label{map_cross}
    \mathcal{E}_{C}(\rho) = \sum_{i=0}^{d-1} \rho_{ii} \sum_{j=0}^{d-1} p^{(i)}_{j} \ketbra{j}{j} + \sum_{i \neq j} [\rho_{ij}(1+\alpha_{ij}) \ketbra{i}{j} + \mbox{h.c.}]~.
\end{equation}

\begin{figure}[h!]
    \centering
    \includegraphics[width=\linewidth]{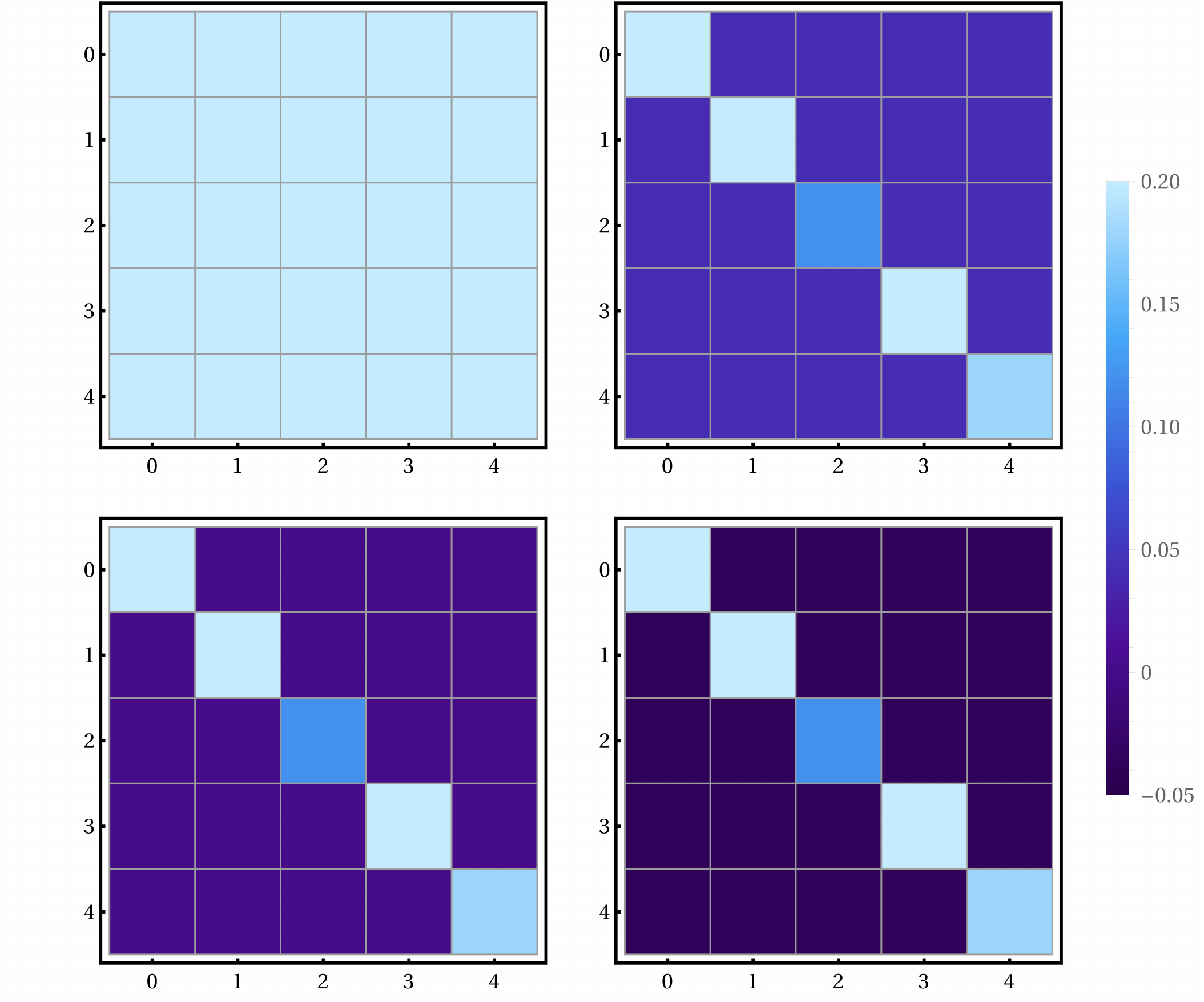}
    \caption{Pictorial representation of the action of $\mathcal{E}_{C}$ on an input state $\rho$ in the core basis $\{\ket{i}\}$, $i=0,\dots,4$. Here, we have assumed $\rho=\frac{1}{5}\sum_{i,j=0}^{4}\ketbra{i}{j}$ and depicted $\mathcal{E}_{C}(\rho)$ for: a) $\alpha=0$ (i.e., the initial state), b) $\alpha = -0.8$, c) $\alpha = -1$, and d) $ \alpha = -1.2$. The crosstalk parameters have been chosen such that $p^{(i)}_{j} = P_{ij}$, where $P$ is the $5 \times  5$ matrix of Eq.(\ref{mat_cross}).}
    \label{fig:mcf}
\end{figure}

\noindent From Eq.(\ref{map_cross}) it is clear that $\mathcal{E}_{C}$ reduces the amplitude of the coherences of an input state, while also scrambling its diagonal terms. As an example, in Fig.\ref{fig:mcf}, we depict the action of $\mathcal{E}_{C}$ on the maximally coherent input state $\rho=\frac{1}{5}\sum_{i,j=0}^{4}\ketbra{i}{j}$ for various values of $\alpha_{ij}=\alpha_{ji} \equiv \alpha \in \mathbb{R}$ and the following choice of the crosstalk parameters, i.e.,
\begin{equation}
\label{mat_cross}
P =\begin{pmatrix*}
0.7 & 0.1 & 0.1 & 0.1 & 0\\
0.2 & 0.5 & 0.1 & 0.1 & 0.1\\
0 & 0.3 & 0.3 & 0.3 & 0.1 \\
0.1 & 0.2 & 0 & 0.6 & 0.1 \\
0.1 & 0.1 & 0.1 & 0.1 & 0.6
\end{pmatrix*}~,
\end{equation}
where $P_{ij}=p^{(i)}_{j}$ for every $i,j \in \{0,\dots, d-1\}$~. Notice that, as an effect of the propagation through the MCF, crosstalk between different cores as well as decoherence effects arise in the input state, causing a scrambling of its diagonal elements and a reduction of the coherences in the core basis.   

\noindent In order to inspect whether the map $\mathcal{E}_{C}$ defines a valid quantum channel we can consider its associated Choi operator $\mathcal{J}_{\mathcal{E}_{C}}$, given by
\begin{align}
\label{choi:crosstalk}
    \mathcal{J}_{\mathcal{E}_{C}} &= \frac{1}{d}\sum_{i,j=0}^{d-1} p^{(i)}_{j} \ketbra{ij}{ij} + \frac{1}{d}\sum_{i \neq j =0}^{d-1} (1+\alpha_{ij}) \ketbra{ii}{jj} ~.
\end{align}
Using Eq.(\ref{tp}) from Theorem \ref{th:choi}, it is easy to see that $\mathcal{E}_{C}$ is trace-preserving if and only if the probabilities $p^{(i)}_{j}$ satisfy 
\begin{equation}
\label{tp_cond}
    \sum_{j=0}^{d-1}p^{(i)}_{j}=1~, \quad  \forall i\in \{0,\dots, d-1\}~.
\end{equation}

\noindent As for complete positivity, let us first notice that the operator $\mathcal{J}_{\mathcal{E}_{C}}$ of Eq.(\ref{choi:crosstalk}) can be cast as
\begin{equation}
\mathcal{J}_{\mathcal{E}_{C}} = \hat{\mathcal{J}}_{\mathcal{E}_{C}}\bigoplus_{i\neq j} p^{(i)}_{j}/d~,
\end{equation}
where $\hat{\mathcal{J}}_{\mathcal{E}_{C}}$ is a $d \times d$ matrix given by
\begin{equation}
    \label{choi:crosstalk:mat}
    \hat{\mathcal{J}}_{\mathcal{E}_{C}} = 
    \frac{1}{d} \begin{pmatrix*}
    p^{(0)}_{0} & 1+\alpha_{01} & \dots & 1 + \alpha_{0, d-1} \\
    1+\alpha^{*}_{01} & p^{(1)}_{1} & \dots & 1 + \alpha_{1,d-1} \\
    \vdots & \vdots  & \ddots & \vdots \\
    1 + \alpha^{*}_{0, d-1} & 1 + \alpha^{*}_{1,d-1}  & \dots & p^{(d-1)}_{d-1} 
    \end{pmatrix*}~.
\end{equation}
Since $p^{(i)}_{j} \geq 0~ \forall i,j$, it is easy to see that $\mathcal{J}_{\mathcal{E}_{C}} \succeq 0 \iff \hat{\mathcal{J}}_{\mathcal{E}_{C}} \succeq 0$. As a consequence, $\mathcal{E}_{C}$ defines a CPTP map iff the set of coefficients $\{\alpha_{ij}\}$ and  probabilities $\{p^{(i)}_{j}\}$ guarantee that $\hat{\mathcal{J}}_{\mathcal{E}_{C}} \succeq 0$, along with the condition $\sum_{j=0}^{d-1}p^{(i)}_{j}=1$, for every $i\in \{0,\dots, d-1\}$.

\subsection{Bound entanglement generation via MCFs}
\label{sec:bound}

\noindent Our proposal for the generation of entangled states using MCFs is based on the simple observation that the Choi operator of the state at the output of a MCF belongs to the class of CLDUI states. Indeed, comparing Eq.(\ref{choi:crosstalk}) to Eq.(\ref{CLDUI}), it is immediate to see that the equivalence is complete setting
\begin{align*}
    \label{condCLDUI}
     &A_{ij} = \frac{p^{(i)}_{j}}{d}~, \forall i,j~, \\
     &B_{ii} =\frac{p^{(i)}_{i}}{d}~, \quad  B_{ij} = \frac{1+\alpha_{ij}}{d}~, ~ \forall i\neq j~,
\end{align*}
\noindent from which it follows that $\rho_{(A,B)} = \mathcal{J}_{\mathcal{E}_{C}}$, with
\begin{equation*}
\label{matrices}
    A = \frac{1}{d} \begin{pmatrix}
       p^{(0)}_{0} 	& p^{(0)}_{1}  & \cdots & p^{(0)}_{d-1} \\
        p^{(1)}_{0}  & p^{(1)}_{1} & \cdots & p^{(1)}_{d-1} \\
       \vdots 	& \vdots & \ddots & \vdots \\
       p^{(d-1)}_{0} & p^{(d-1)}_{1} & \cdots & p^{(d-1)}_{d-1} 
    \end{pmatrix}~, \quad B = \hat{\mathcal{J}}_{\mathcal{E}_{C}}~.
\end{equation*}

\noindent This correspondence is extremely interesting, since it suggests a method to practically implement CLDUI states. In fact, as a consequence of the Choi-Jamio\l kowski isomorphism, the desired state can be generated starting from a maximally entangled state  $\ket{\Psi^{+}} = \frac{1}{\sqrt{d}} \sum_{i=0}^{d-1} \ket{i i}$ and sending one party through the MCF, while leaving the other unaffected, i.e., $\rho_{(A,B)} = (\mathds{1} \otimes \mathcal{E}_{C})\left( \ketbra{\Psi^{+}}{\Psi^{+}}\right)$. 
Moreover, the entanglement in the output state can be certified by checking whether the matrices $A$ and $B$ of Eq.(\ref{matrices}) violates the condition of Theorem \ref{th:realign_AB}. Notice that this task amounts to measure the crosstalk parameters and the off-diagonal density matrix elements (performing, e.g., quantum tomography), thus representing an experimentally accessible criterion. Remarkably, our technique allows to generate not only NPT states, but also families of bound entangled DS states. 
Indeed, assuming $\alpha_{ij}\in \mathbb{R}$ and choosing the crosstalk parameters $p^{(i)}_{j}$ such that
\begin{align*}
    \label{cond}
    &p_{ii} = \frac{p^{(i)}_{i}}{d} ~, \quad   &\frac{p_{ij}}{2} = \frac{p^{(i)}_{j}}{d} = \frac{p^{(j)}_{i}}{d} = \frac{1+\alpha_{ij}}{d}~,   \quad \forall i\neq j~,  
\end{align*}
\noindent it is easy to see that $\mathcal{J}_{\mathcal{E}_{C}} = \rho_{DS}^{\Gamma}$, with $ \hat{\mathcal{J}}_{\mathcal{E}_{C}} = M_{d}(\rho_{DS})$. \noindent Hence, recalling Theorem \ref{th:CAR}, the certification of bound entangled state can be tackled checking whether the matrix $M_{d}(\rho_{DS})$ is such that $M_{d}(\rho_{DS}) \in \mathcal{DNN}_{d}\setminus \mathcal{CP}_{d}$. 
The only caveat is that the explicit expression of the state $\mathcal{J}_{\mathcal{E}_{C}}$ depends on a set of parameters that are the result of two detrimental phenomena, namely the crosstalk among different cores and the dephasing induced by the propagation through the quantum channel. For this reason, in the context of a practical implementation of this protocol, it is essential that the experimenter is able to carefully manipulate such sources of noise. Moreover, in order to guarantee that a bound entangled state $\mathcal{J}_{\mathcal{E}_{C}}$ can be created deterministically through our protocol, it is necessary that such state is the result of a trace-preserving quantum channel $\mathcal{E}_{C}$. Recalling Eqs.(\ref{tp_cond}), this latter requirement imposes an extra constraint on $M_{d}(\rho_{DS})$, i.e.,
\begin{equation}
\label{mat_cond}
    \sum_{i} M_{d}(\rho_{DS})_{ij}= \sum_{j} M_{d}(\rho_{DS})_{ij} =\frac{1}{d}~, \quad \forall i,j.
\end{equation}

\noindent We conclude this section providing an explicit example of a $6 \otimes 6$ PPT-entangled DS state whose associated matrix is such that $M_{6}(\rho_{DS}) \in \mathcal{DNN}_{6} \setminus \mathcal{CP}_{6}$ and satisfies Eq.(\ref{mat_cond}), i.e., \cite{sonneveld2009nonnegative,tura2018separability} 
\begin{equation*}
    M_{d}(\rho_{DS}) =
    \frac{1}{6}\begin{pmatrix*}
    1/3 & 1/4 & 1/12 & 0 & 1/12 & 1/4\\
    1/4 & 1/3 & 1/4 & 1/12 & 0 & 1/12\\
    1/12 & 1/4 & 1/3 & 1/4 & 1/12 & 0\\
    0 & 1/12 & 1/4 & 1/3 & 1/4 & 1/12\\
    1/12 & 0 & 1/12 & 1/4 & 1/3 & 1/4\\
    1/4 & 1/12 & 0 & 1/12 & 1/4 & 1/3
    \end{pmatrix*}~.
\end{equation*}
~

\section{Conclusion}
\label{sec:conc}

\noindent In this work, we have investigated the role of multicore fibres as a tool to manipulate entanglement in quantum states. To this end, we have modeled the action of a multicore fibre on an input state using the formalism of quantum channels. Leveraging this insight, we have proposed a theoretical scheme where,
starting from a maximally entangled state of two qudits, it is possible to generate new classes of states whose entanglement properties can be certified by measuring the crosstalk parameters of the multicore fibre. Remarkably, for a suitable choice of such parameters, it is also possible to retrieve a whole family of bound entangled states at the output of the MCF. We believe our work opens new avenues for the practical realization of bound entangled states, paving the way for the design of enhanced quantum networks.

\section{Acknowledgements}
This research has been co-funded by the European Union - NextGeneration EU, "Integrated infrastructure initiative in Photonic and Quantum Sciences" - I-PHOQS [IR0000016, ID D2B8D520, CUP B53C22001750006],by the European Union ERC StG, QOMUNE, 101077917, by the Project SERICS (PE00000014) and STARTrAP (P2022LB47E) under the MUR National Recovery and Resilience Plan funded by the European Union - NextGenerationEU, by the Project QuONTENT under the Progetti di Ricerca, CNR program funded by the Consiglio Nazionale delle Ricerche (CNR) and by the European Union - PON Ricerca e Innovazione 2014-2020 FESR - Project ARS01/00734 QUANCOM.

\bibliographystyle{apsrev4-1}
\bibliography{biblio}

\end{document}